\documentclass[aip,amssymb,amsmath,apl,reprint,noshowpacs]{revtex4-1}
\usepackage{times}
\usepackage{amssymb,amsmath,graphicx}
\usepackage[T1]{fontenc}
\usepackage[utf8]{inputenc}
\usepackage[english]{babel}

\usepackage{textcomp}

\newcommand{\figwidth}{0.95\columnwidth} 
\newcommand{\commentOut}[1]{}
\usepackage{scalefnt}   

\newcommand{\abs}[1]{\left| #1 \right|} 

\renewcommand{\d}[2]{\frac{\text{d} #1}{\text{d} #2}} 

\newcommand{\affil}{Photonics Laboratory, ETH Zürich, CH-8093 Zürich, Switzerland}
\newcommand{\affilTwo}{Department of Physics, ETH Zürich, CH-8093 Zürich, Switzerland}

\begin{document}
\title{The role of titanium in electromigrated tunnel junctions}
\author{Martin Frimmer}
\affiliation{\affil}
\email{frimmerm@ethz.ch}
\homepage{http://www.photonics.ethz.ch}
\author{Gabriel Puebla-Hellmann}
\affiliation{\affilTwo}
\author{Andreas Wallraff}
\affiliation{\affilTwo}
\author{Lukas Novotny}
\affiliation{\affil}

\begin{abstract}
A standard route for fabrication of nanoscopic tunnel junctions is via electromigration of lithographically prepared gold nanowires. In the lithography process, a thin adhesion layer, typically titanium, is used to promote the adhesion of the gold nanowires to the substrate. Here, we demonstrate that such an adhesion layer plays a vital role in the electrical transport behavior of electromigrated tunnel junctions. We show that junctions fabricated from gold deposited on top of a titanium adhesion layer are electrically stable at ambient conditions, in contrast to gold junctions without a titanium adhesion layer. We furthermore find that electromigrated junctions fabricated from pure titanium are electrically exceptionally stable. Based on our transport data, we provide evidence that the barrier in gold-on-titanium tunnel devices is formed by the native oxide of titanium.
\end{abstract}
\date\today

\maketitle

\section{Introduction}
Nanoscience is continuously pushing the dimensions of electronic building blocks to ever smaller sizes, currently approaching the fundamental limit of the size of a single atom.\cite{Scheer2010} On the one hand, this size reduction poses a significant challenge to our understanding, since concepts and insights carried over from bulk systems are expected to reach their limits in systems whose characteristic size approaches that of a single atom. On the other hand, we expect new phenomena to emerge at such reduced length scales, possibly offering room for new functionalities.
One device of current interest is the atomic tunnel junction, where transport between two conductors happens through an insulating barrier across a few or (ideally) even a single atom on either side of the barrier. The nonlinear electronic transport characteristics of tunnel junctions have attracted great interest in the past to mix and rectify AC electric fields.\cite{Faris1973,Elchinger1976,Heiblum1978,Gutjahr-Loeser1999,Tu2006}  Recently, in the wake of plasmonics,\cite{Gramotnev2010} the interest in tunnel junctions has been revived in the context of light generation by inelastic tunneling\cite{Bharadwaj2011} and light detection by optical rectification.\cite{Ward2010} A key requirement to elucidate the physics of electronic transport across atomic-sized tunnel junctions are stable devices of well defined electrode and barrier materials.\cite{Prins2009,Mangin2009a} Most experiments have been conducted with Au as an electrode material. Au does not oxidize at ambient conditions and can be readily functionalized, a fact relished by the molecular electronics community.\cite{Scheer2010} However, Au electrodes are unstable at room temperature.\cite{Prins2009}
The most popular fabrication method for tunnel junctions has been electromigration of Au wires of diameters of the order of 100\,nm.\cite{Park1999,Strachan2005} Such metallic wires are usually fabricated by lithographically patterning a resist and subsequent physical vapor deposition of the metal followed by lift off. Due to the poor adhesion of Au to SiO$_2$, the insulating substrate of choice, a thin adhesion layer between the Au layer and the substrate is commonly used.\cite{Ward2010,Stolz2014} Typically, Ti or Cr are chosen as adhesion layers, where the adhesion layer thickness is about 1--10\% of the supported Au layer thickness.
Surprisingly, it is typically assumed that tunnel junctions fabricated from electromigrated Au wires supported by an adhesion layer form Au-air-Au tunnel junctions and the presence of the adhesion layer has never been taken into account.\cite{Ward2010,Stolz2014} It is therefore an open question if the adhesion layer influences the electrical transport behavior of electromigrated Au-based tunnel junctions.

In this paper, we study the role of a Ti adhesion layer in the electromigration process and the transport behavior of a Au wire it supports at ambient conditions. We find that the Ti layer stabilizes the electrical transport behavior of the junction. Tunnel devices fabricated from Au nanowires by electromigration without the use of an adhesion layer exhibit highly unstable current-voltage characteristics. Furthermore, we demonstrate that electromigration of pure Ti nanowires leads to the formation of highly reproducible and exceptionally stable tunnel junctions. Our results show that the Ti adhesion layer significantly influences the electronic transport behavior of electromigrated tunnel junctions.
In particular, our transport data suggest that the tunnel barrier in electromigrated nanogaps in Au wires supported by a Ti adhesion layer is formed by the native oxide TiO$_2$.

\section{Sample fabrication}
Our devices are fabricated on Si wafers covered by 280\,nm of thermal oxide. We use standard optical lithography to generate macroscopic bonding pads to electrically contact the structures. Subsequently, we pattern a polymethyl-methacrylate (PMMA) positive resist layer with electron beam lithography (EBL) to generate a nanowire that is nominally 400\,nm long and 150\,nm wide. After (electron beam) evaporation of the metal the resist layer is lifted off. All samples are cleaned in an O$_2$ plasma in a reactive ion etcher to remove any possible organic contamination.

We mount our samples on a chip carrier, contact each nanowire by wire bonding and use a source meter (Keithley 2602B) to bias the nanowires. Our devices show a resistance of about 500\,$\Omega$ prior to electromigration. This resistance is dominated by the mm-length leads connecting the nanowire with the bonding pads. For controlled electromigration,\cite{Campbell2013} we ramp up the bias voltage $V$ at a rate of 100\,mV/s while monitoring the current $I$. When the derivative $\d{I}{V}$ drops below a certain threshold value, the bias is set to zero and a new voltage ramp is started. We repeat this process until the resistance of the junction is of the order of 100\,k$\Omega$ or higher.

\section{Ti/Au junctions}
\begin{figure}
\includegraphics[width=\figwidth]{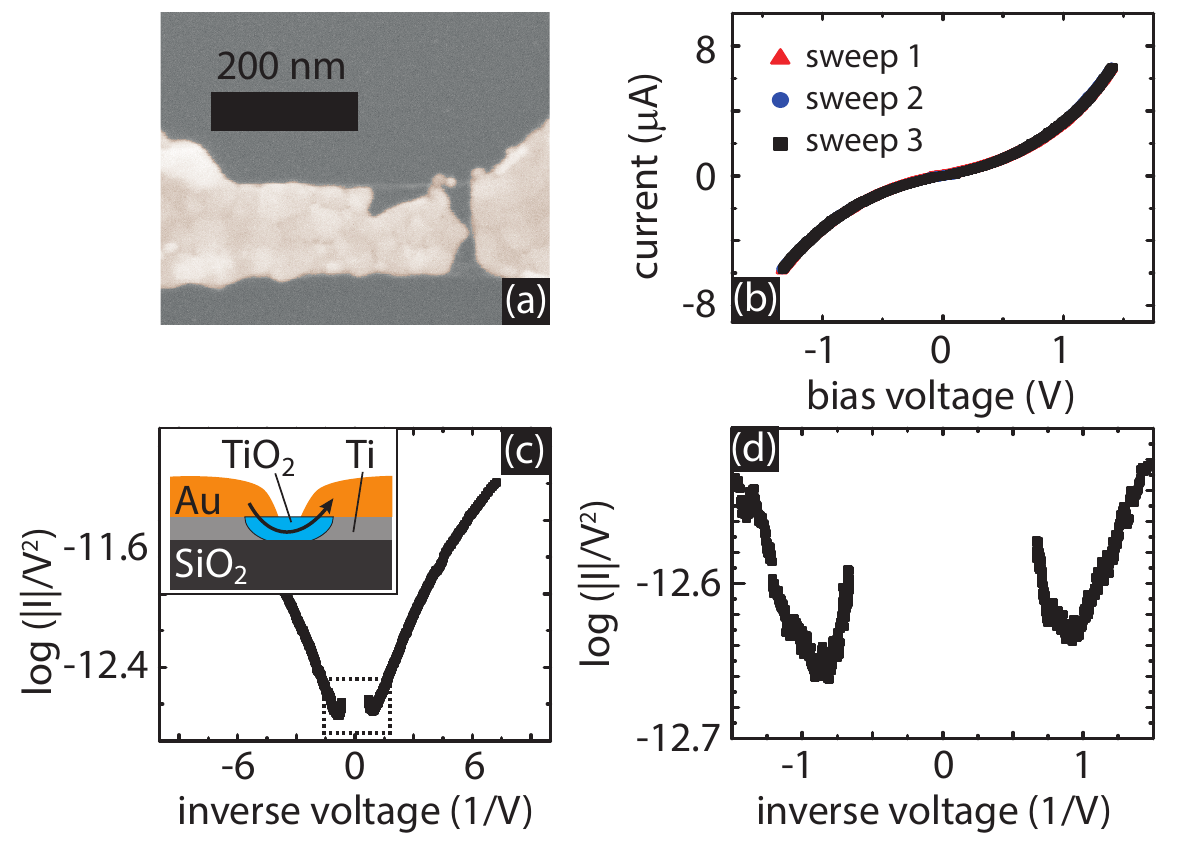}
\caption{(a)~Top view SEM image (false colored) of nanowire fabricated by evaporation of 35\,nm Au on top of a 3\,nm Ti adhesion layer after electromigration. (b)~Three IV-curves of junction imaged in (a). The junction is in the tunneling regime, exhibiting a nonlinear dependence of current on voltage. Note that the three voltage sweeps are identical to within less than the symbol size. (c)~Fowler-Nordheim plot of IV curve in (b). Inset: Sketch of junction region after electromigration. Formation of a gap in the gold layer exposes the Ti adhesion layer which oxidizes and forms a path for the tunneling electrons. (d)~Magnified view of the region marked with a box in (c). Inflection points appear at $\pm$0.9~V$^{-1}$.}
\label{fig:TiAuJunction}
\end{figure}
We have fabricated nanowires by evaporation of a 35\,nm Au layer onto a 3\,nm Ti adhesion layer. In the following, we refer to this layer arrangement as Ti/Au. We have investigated twelve such Ti/Au junctions and have observed the behavior reported below in ten of those devices. The remaining two have been destroyed by electrostatic discharge, a frequent problem in the fabrication and handling of tunneling devices.\cite{Puebla-Hellmann2012}  The scanning electron microscopy (SEM) image in Fig.~\ref{fig:TiAuJunction}(a) shows a typical functional Ti/Au junction after electromigration. We clearly observe a gap formed at the right end of the nanowire. Typical current-voltage (IV) curves of such a Ti/Au tunnel junction after electromigration are shown in Fig.~\ref{fig:TiAuJunction}(b). Note that three consecutive IV-sweeps are plotted, which are identical to within the symbol size. Accordingly, Ti/Au junctions are highly stable even at room temperature. The non-linear characteristic of the IV-curve of the Ti/Au junction illustrates that the junction is in the tunneling regime.\cite{Simmons1963} In a Fowler-Nordheim (FN) representation [see Figs.~\ref{fig:TiAuJunction}(c) and (d)],\cite{Fowler1928} where $\text{log}(\abs{I}/V^2)$ is plotted against $1/V$, we find inflection points at $\pm$0.9\,V$^{-1}$, which suggest a tunnel barrier height of 1.1\,eV.\cite{Beebe2006} We observe that the value of the inflection point in the FN plot varies from device to device in a range of about 0.6\,eV to 1.5\,eV, in accordance with previous reports.\cite{Mangin2009a} Both the fact that the barrier height varies from device to device and the fact that the barrier height is typically far from the work function of Au ($W_\text{Au}=5.1\,\text{eV}$\cite{CRCHandbook2007}), which would be the barrier height expected from an ideal Au-air-Au junction, have been attributed to a work function modification at atomically sized nanogaps\cite{Ward2010} or to surface contamination.~\cite{Mangin2009a}

However, for a tunnel junction supported by a dielectric substrate the increased field strength in the dielectric suggests a tunneling path through the substrate instead of through air.
We note that such a tunneling path through the supporting layer can explain our experimentally determined barrier height of $\approx1$\,eV when taking into account that Ti forms a natural oxide at ambient conditions. From the SEM image in Fig.~\ref{fig:TiAuJunction}(a) it is obvious that Au forms a constriction on the wire during electromigration thereby exposing the Ti adhesion layer which subsequently oxidizes. Accordingly, it is possible that the transport through the junction is through a Au/TiO$_2$/Au barrier as illustrated in the inset of Fig.~\ref{fig:TiAuJunction}(c). In fact, the difference in work functions of Au ($W_\text{Au}=5.1$\,eV\cite{CRCHandbook2007}) and the electron affinity of TiO$_2$ ($\phi_\text{TiO$_2$}=4.0$\,eV\cite{Hope1983}) indeed suggest a barrier height of 1.1\,eV, assuming the barrier height is given by $W_\text{Au}-\phi_\text{TiO$_2$}$.\cite{Tung1993} For the Au/TiO$_2$ interface a Schottky barrier height of 0.9\,V has been reported,\cite{McFarland2003} which is close to our experimentally determined values for the tunnel barrier height.

\section{Pure Au junctions}
\begin{figure}
\includegraphics[width=\figwidth]{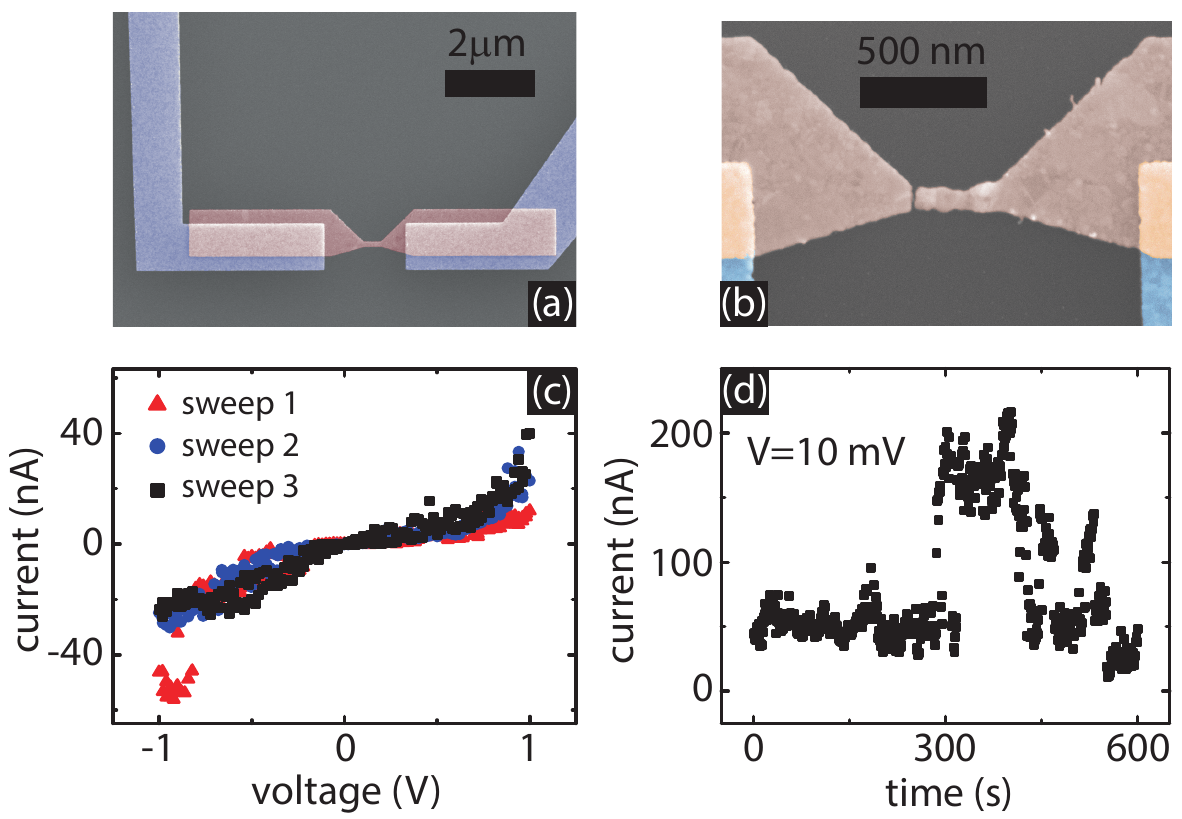}
\caption{(a)~SEM image (false colored) of the fabricated pure Au junctions supported by an SiO$_2$ substrate. Large contact pads of Au (blue) were deposited on top of a Ti adhesion layer leaving a 2~µm gap. In a second step, a pure Au layer (red, no adhesion layer) is deposited to bridge the gap with a nanowire. (b)~SEM image (false colored) of junction after electromigration showing that a tunneling gap has formed in the pure Au nanowire. (c)~IV-curves of pure Au junction. The transport behavior differs from sweep to sweep and also during a single sweep, demonstrating the instability of pure Au junctions. (d)~Current across a pure Au junction monitored over 600\,s under a constant bias voltage $V=10$\,mV. The current varies over more than one order of magnitude.}
\label{fig:pureAuJunction}
\end{figure}
Due to the poor adhesion of Au to SiO$_2$ it is difficult to avoid the use of an adhesion layer. Approaches towards adhesion layer-free fabrication of Au nanowires for electromigration have used shadow evaporation under an angle, which creates very short wires without adhesion layer.\cite{Park1999,Mangin2009a} Since the wire will not necessarily break at its thinnest point [in our case the wire always breaks where it is connected to the trapezoidal contacts, see Figs.~\ref{fig:TiAuJunction}(a), \ref{fig:pureAuJunction}(b)], a sufficiently long wire without adhesion layer has to be fabricated to rule out the presence of the adhesion material at the tunneling gap. We use a double step EBL process to create adhesion-layer free Au nanowires on a SiO$_2$ substrate [see SEM image in Fig.~\ref{fig:pureAuJunction}(a)]. After fabrication of the macroscopic contact pads by optical lithography [not shown in Fig.~\ref{fig:pureAuJunction}(a)] we use EBL to create mesoscopic contacts [blue in Fig.~\ref{fig:pureAuJunction}(a)] leaving a gap of a few $\mu$m. These mesoscopic contacts are made of 35\,nm of Au on top of a 3\,nm Ti adhesion layer. In a second EBL step, we define the nanowire in a bridge [red in Fig.~\ref{fig:pureAuJunction}(a)] connecting the mesoscopic contacts created in the first EBL step. The bridge is formed by evaporation of 35\,nm of Au \emph{without} the use of an adhesion layer. The adhesion of the pure Au bridge to the mesoscopic Ti/Au contacts is sufficient to keep it in place.
We have fabricated 12 devices, out of which 2 were destroyed by electrostatic discharge. In the following, we report the typical behavior observed on the remaining devices.
While the electromigration of the pure Au junctions proceeds as that of the Ti/Au junctions throughout the $\Omega$ and k$\Omega$ regime, the pure Au junctions become highly unstable in the 100~k$\Omega$ range. In Fig.~\ref{fig:pureAuJunction}(c) we show three consecutively taken IV-curves of a pure Au junction which show the instability and random switching of the conduction behavior. The random switching does not only occur at large bias voltages, as is shown in Fig.~\ref{fig:pureAuJunction}(d), where we plot the current of a typical pure Au junction over 10 minutes under a constant bias of 10\,mV. The current varies drastically by more than one order of magnitude. It is well known that Au atoms are very mobile both on the surface of bulk Au and on SiO$_2$ at room temperature.\cite{Prins2009} This renders pure Au junctions unsuitable for the fabrication of stable tunneling junctions at ambient conditions.

\section{Pure Ti junctions}
\begin{figure}
\includegraphics[width=\figwidth]{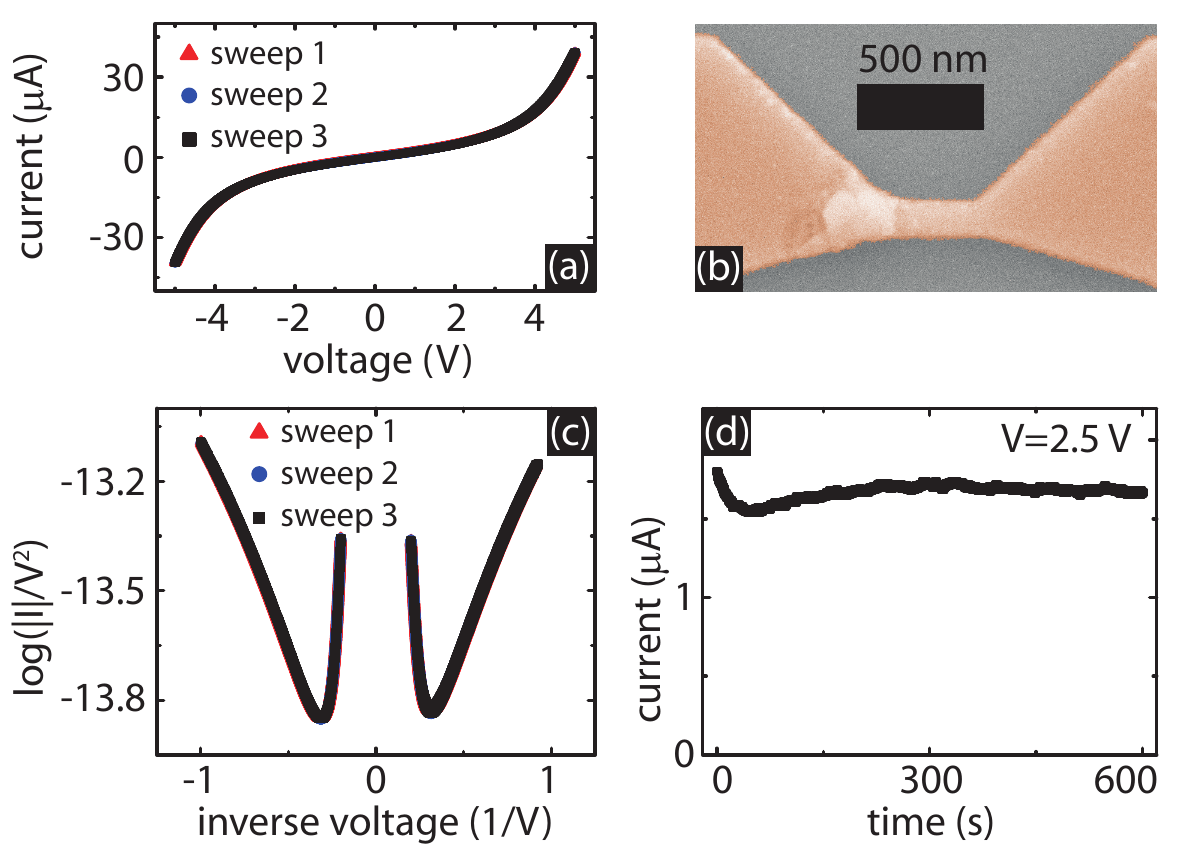}
\caption{(a)~IV-curves of pure Ti junction. Note that three consecutive measurements are plotted which lie on top of each other, illustrating the stability of the junction. (b)~SEM image (false colored) of the pure Ti junction after electromigration, showing an irregularity at the left end of the nanowire. (c)~Fowler-Nordheim plot of the data in (a) showing inflection points at $\pm0.30$\,V$^{-1}$. (d)~Current through pure Ti junction monitored over 600\,s at a bias voltage of 2.5\,V which was instantaneously turned on at $t=0$. After a short transient period the current is stable to within a few per cent.}
\label{fig:pureTiJunction}
\end{figure}
Since we found unstable transport properties for pure Au junctions but stable transport properties for Ti/Au junctions, it is reasonable to conjecture that the Ti adhesion layer indeed has a significant impact on the transport behavior of Ti/Au junctions. We therefore fabricated a sample of pure Ti nanowires of 35\,nm thickness which we electromigrated. We note that controlled electromigration of Ti has not been reported to date. It turns out that the electromigration of pure Ti nanowires is extremely well behaved and the resulting tunnel junctions are highly reproducible and stable. Five out of five fabricated devices were functional and yielded the results reported in the following. In Fig.~\ref{fig:pureTiJunction}(a) we show three consecutive IV-sweeps of an electromigrated Ti junction. The three curves lie perfectly on top of each other and the error bars on the measurements are accordingly smaller than the symbol size. We stress that the junctions are stable to remarkably high bias voltages of $\pm$5\,V. To further illustrate the stability of these pure Ti junctions we plot the current at a large applied bias of 2.5\,V over a period of 10 minutes in Fig.~\ref{fig:pureTiJunction}(d). After a short transient period, the current is stable to within a few percent. When plotting the IV-curve from Fig.~\ref{fig:pureTiJunction}(a) in a Fowler-Nordheim representation in Fig.~\ref{fig:pureTiJunction}(c) we find inflection points at approximately $\pm0.30$\,V$^{-1}$, corresponding to a barrier height of 3.3\,eV. Inspection of the electromigrated pure Ti-junction in the SEM image in Fig.~\ref{fig:pureTiJunction}(b) shows a significantly different picture compared to the Ti/Au and pure Au junctions. While on pure Au and Ti/Au junctions a narrow constriction has formed during the migration process [compare Figs.~\ref{fig:TiAuJunction}(a) and \ref{fig:pureAuJunction}(b)], the pure Ti junction rather forms a region where the surface appears irregular and rough, as seen at the left end of the nanowire in the SEM image in Fig.~\ref{fig:pureTiJunction}(b). This irregularity has not been present prior to electromigration. Such physical deformation of electrodes due to an applied bias voltage has been reported previously for tunnel devices involving Ti and its oxide and has been attributed to bubble formation as a result of the drift of oxygen ions in TiO$_2$.\cite{Huang2010}

Since the transport behavior is clearly dominated by a tunneling process, but no observable gap or constriction has formed in Fig.~\ref{fig:pureTiJunction}(b), the question is what constitutes the tunneling barrier in the pure Ti junctions. Since Ti forms a native stochiometric oxide when in contact with oxygen,\cite{Fracassi1992} it is conceivable that during electromigration the Ti wire is increasingly thinned down until the oxide pinches off the conducting path and a tunnel junction is formed. Accordingly, we would expect that the transport behavior of the electromigrated pure Ti junctions is described by transport across a Ti/TiO$_2$/Ti junction. However, the large barrier of 3.3\,eV observed for the tunnel junctions fabricated from pure Ti cannot be explained by considering an ideal Ti/TiO$_2$/Ti junction, since the electron affinity $\phi_\text{TiO$_2$}=4.0$\,eV of TiO$_2$ is very close to the work function of Ti ($W_\text{Ti}=4.3$\,eV\cite{CRCHandbook2007}), which would result in a barrier height $W_\text{Ti}-\phi_\text{TiO$_2$}=0.3$\,eV. Keeping in mind that TiO$_2$ is a wide-bandgap semiconductor, it is well known that defects and impurities can render the tunnel barrier heights at Schottky contacts largely determined by the fabrication process.\cite{Tung1993} Previous experimental studies on Ti/TiO$_2$ contacts have reported a Schottky barrier of 0.13\,eV\cite{Huang2010} or Ohmic behavior.\cite{McFarland2003}
We note that our experimentally found value for the tunnel barrier height in electromigrated pure Ti junctions of 3.3\,eV is close to the band gap of TiO$_2$, which is about 3.2\,eV.\cite{Morikawa2001}

\section{Discussion and conclusion}
In our experiments, junctions created by electromigration from pure Au nanowires have yielded unstable tunnel junctions. On the other hand, junctions fabricated from pure Ti or containing Ti as an adhesion layer under a much thicker Au layer were found to be highly stable. For the stable junctions we have found tunnel barrier heights that are incommensurate with a metal/air/metal junction assuming bulk values for work functions and electron affinities, but can be explained by assuming a metal/TiO$_2$/metal barrier. Clearly, in tunnel junctions fabricated with the use of a Ti adhesion layer, this layer plays a vital role in the transport behavior of the junction. We point out that TiO$_2$ is a photoconductive semiconductor.\cite{Haga1997,Tsukamoto2013} Accordingly, despite the promising stability observed in our electromigrated Au wires supported by a Ti adhesion layer and our tunnel junctions fabricated from pure Ti, further work is necessary to clarify potential applications of Ti-based tunnel junctions, in particular regarding their suitability for studying light induced transport.

\begin{acknowledgments}
This work is supported by the Swiss National Science Foundation (SNF) under grant 200021-146358.
\end{acknowledgments}

\bibliography{Bibliography_Frimmer}

\end{document}